\documentstyle[12pt,psfig]{article}
\def\frac#1#2{{\textstyle{{#1}\over {#2}}}}

\def\lsim{\mathrel{\rlap{\lower4pt\hbox{\hskip1pt$\sim$}}
    \raise1pt\hbox{$<$}}}
\def\gsim{\mathrel{\rlap{\lower4pt\hbox{\hskip1pt$\sim$}}
    \raise1pt\hbox{$>$}}}
\def\sqr#1#2{{\vcenter{\vbox{\hrule height.#2pt
         \hbox{\vrule width.#2pt height#1pt \kern#1pt
         \vrule width.#2pt}
         \hrule height.#2pt}}}}

\newcommand{\beq}{\begin{equation}}
\newcommand{\eeq}{\end{equation}}
\newcommand{\bea}{\begin{eqnarray}}
\newcommand{\eea}{\end{eqnarray}}

\renewenvironment{thebibliography}[1]
 { \rm
   \begin{list}{\arabic{enumi}.}
    {\usecounter{enumi} \setlength{\parsep}{0pt}
     \setlength{\itemsep}{3pt} \settowidth{\labelwidth}{#1.}
     \sloppy
    }}{\end{list}}
 
\begin{document}
\titlepage
 
\begin{flushright}
{IUNTC 01-01\\}
{May 2001\\}
\end{flushright}

\vglue 1cm

\begin{center}
{{\bf ROLE OF PHOTOPRODUCTION IN EXOTIC MESON SEARCHES }
\vglue 1.0cm
{Adam P. Szczepaniak and Maciej Swat\\} 

{\it Physics Department and Nuclear Theory Center, \\}
{\it Indiana University, Bloomington, IN 47405, U.S.A.\\}
}
\vglue 0.8cm
  
\end{center}
 
{\rightskip=2pc\leftskip=2pc\noindent
We discuss two production mechanisms of the $J^{PC}=1^{-+}$ 
 exotic meson,  hadroproduction, using pion beams and photoproduction. 
 We show that the 
  ratio of exotic to non-exotic, in particular the $a_2$,  meson production 
 cross sections  is expected to be by a factor of 5 to 10 larger, 
 in photoproduction then in hadroproduction. Furthermore we show 
 that the low-t photoproduction  of exotic meson is enhanced as compared to 
 hadronic production. 
 This findings support the simple quark 
 picture in which exotic meson production is predicted to be 
 enhanced when the beam is a virtual $Q\bar Q$ pair with a spin-1
 (photon) rather then  with a spin-0 (pion). 
 
}

\vskip 1 cm

\newpage
 
\baselineskip=20pt

{\it 1. Introduction.} 
  Present understanding of confinement is still mostly
  qualitative. Even though in recent years significant progress has
  been made in lattice gauge studies of QCD,  
   the dynamics of confinement, especially in 
  association with light quarks, remains mysterious. Similarly, 
 it remains to be verified whether soft gluonic excitations responsible for
 confinement are present in the low energy 
  resonance spectrum. 
 
 In the past few years, a number of strong candidates for states 
   lying outside the valence quark model spectrum have been 
  reported. Examination of the scalar-isoscalar mesons produced in
  $p\bar p$ annihilation~\cite{CB}, central production and
  $J/\psi$~\cite{CP} radiative
  decays~\cite{JP} indicates that in the
   $1-2\mbox{ GeV}$ mass range there is 
  an overpopulation of states as compared to predictions 
 of the valence quark model. 
 Furthermore analysis of the decay patterns seems to indicate that 
 some of these states may have a significant non-$Q\bar Q$ component. 
 Since these states can mix with the regular $Q\bar Q$ mesons 
 the non-$Q\bar Q$ components cannot be disentangled in a model
   independent way.  A much cleaner signatures for non-$Q\bar Q$
   and/or gluonic excitations could come from exotic states. 
 By definition, these are states 
 which have combinations of  spin, parity and charge conjugation quantum
 numbers that cannot be attributed to the valence quark degrees of freedom, 
{\it e.g.}  $J^{PC} = 0^{--}, 0^{+-}, 1^{-+}, \cdots $. Recently, the 
 E852 BNL collaboration  
 has reported an exotic, resonant signal with $J^{PC}=1^{-+}$ 
 in the reaction $\pi^- p \to X p$ at $18\mbox{ GeV}$ 
 in two decay channels, $X\to \rho\pi$~\cite{rpi} 
 and $\eta'\pi$~\cite{etappi} corresponding to the resonance mass of 
 $M_{X} \approx 1.6\mbox{ GeV}$ and width of 
  $\Gamma \approx 170\mbox{ MeV}$ and $\Gamma \approx 340\mbox{ MeV}$ 
 in the two
 channels respectively. 
 The existence of the $J^{PC}=1^{-+}$ exotic meson below $2\mbox{ GeV}$ has
 been reported elsewhere~\cite{1mp}. Furthermore, lattice and model
   calculations indicate that the lightest exotic meson indeed should
   have $J^{PC}=1^{-+}$ and mass below 
$2\mbox{ GeV}$~\cite{latt,adia,mod}.

The exotic resonance measured by the E852 collaboration is observed to
 be produced predominantly
 via natural parity exchange, and so is the nearby
 $a_2(1320)$ resonance. The later is a well established regular $Q\bar Q$
 state. Furthermore, in the $\rho\pi$ decay channel the exotic wave 
 is observed to be roughly $5\%$ of the $a_2$ wave. In the $\eta' \pi$
 channel the exotic signal seems to be much stronger but it is also much
 broader. 

It has been argued that real photons may be a better
  source of exotic mesons in high-$s$, low-$t$ reactions~\cite{Isgur}. 
 The argument is based on a simple quark
 picture. At high energies photon couples to the $Q\bar Q$ pair in
 total helicity-1 state, corresponding (in the $Q\bar Q$ rest frame) 
  to the quark-antiquark pair being in the spin-1 configuration. The
  majority of  models dealing with soft gluons predict that the low lying 
 gluonic excitations 
  are in the $TE$ mode {\it i.e.} with 
 $J_g^{PC} = 1^{+-}$~\cite{mod,mod2}. This is also supported by the recent
  lattice calculations of the excited adiabatic potentials for heavy
  quarkonia~\cite{adia}. 
  As a consequence, the 
  $1^{-+}$ ground state has the $Q\bar Q$ pair in $S=1$, 
 so that  $J^{PC}_{Q\bar Q} \times J^{PC}_{g} = 1^{--} \times
 1^{+-} = 1^{-+}$. Thus, according to this simple quark picture 
one would expect 
 exotic mesons  production to be enhanced in reactions with real
  photon ($Q\bar Q$ in spin-1) 
   as compared to reactions with pseudoscalar ($\pi$,
 $K$) meson beams. 

 In this letter  we show that this microscopic picture is 
  supported by the standard, Regge phenomenology of high-s, low-t
  peripheral production. In particular we show that exotic meson
  production is expected to be comparable to production of other, 
non-exotic resonances in particular the $a_2$ meson, and to be enhanced at
  low-t as compared to hadronic production with pion beams.

{\it 2. Peripheral meson production.} 
We begin with the analysis of the reaction $\pi N \to X N \to M_1 \cdots
M_n N $ at high-s and low-t. 
 In particular, a good candidate decay channel  for 
exotic meson searches is $M_1\cdots M_n = (3\pi)^\pm$. The charged
$3\pi$ system has $I=1$ and $G=-1$, thus assuming isospin
 invariance a resonating $3\pi$ $P$-wave belongs to the exotic $1^{-+}$
 isovector multiplet (a neutral $3\pi$ system will overlap with
  the isospin-0, {\it i.e.} non-exotic, $1^{--}$, $\omega$-like
  resonance).  The E852 data corresponds 
  to $\pi^- p \to \pi^+ \pi^- \pi^- p$ at the lab energy 
 $p_L=18\mbox{ GeV}$~\cite{rpi}. The $3\pi$ mass  spectrum is dominated by 
 the $a_2(1320)$ $J^{PC}=2^{++}$ resonance  and at higher mass, 
 the $\pi_2(1600)$ with $J^{PC}=2^{-+}$ is seen together
  with a small $\pi_1(1600)$, $J^{PC}=1^{-+}$  exotic signal. 
 Both the $a_2$ and the exotic, $\pi_1$ resonances are observed to be 
 produced via natural parity exchange and appear in the $D_+ = 1/\sqrt{2}
(D_{M=1} + D_{M=-1})$ and $P_+$ waves respectively. Here
$D_M,P_M,{\it etc.}$
 refer to the t-channel (Gottfried-Jackson) amplitudes corresponding
 to the spin quantization axis along the direction of the beam in the
  resonance rest frame. 
 In the following we will make a simplifying assumption that the natural 
 parity exchange can be saturated by the 
 $\rho$-trajectory. This is certainly a good assumption 
 for the case of a charge exchange reaction {\it e.g.} 
$\pi^- p \to \pi^+\pi^- \pi^0 n$~\cite{a2pionold1}. In the neutral case 
  $\rho$-exchange should be supplemented by the $f_0$ and/or Pomeron 
 exchange contributions~\cite{a2pionold2}. In the later case,
  which, corresponds to the E852 measurement, the 
 estimates for the exotic meson natural exchange amplitude should be
 considered as an upper bound. The $\rho$ and scalar
 meson exchange contribute mainly to the nucleon helicity-flip and
 -nonflip respectively {\it i.e.} are essentially non-interfering. 
 The E852 data lacks absolute normalization therefore we can only use
 it to 
  establish the relative strength of the exotic to $a_2$ production
 yields. We will compare our predictions for the charge exchange 
 $a_2$ production with the data from 
 Ref.~\cite{a2pionold1}.

The $t$-channel amplitude for the reaction $\pi^- p \to X^0 n$ 
 producing a natural parity, $J^{PC}$ resonance via $\rho$-exchange 
 can be written as $A=A(s,t,\lambda'N,\lambda_N,\lambda)$ (neglecting
 the pion mass),

\begin{equation}
A_\pi = g_{X\rho\pi} \sqrt{ {{2J + 1}\over {4\pi}} }
\sqrt{{{-t'}\over {4m^2_N}}} R_\rho(t,s) e^{b_\rho t/2}
 \left( {{m_X^2 - t} \over {2m^2_X}}\right)^{J-1}
 \left[
 G_V I + \sqrt{{{-t'}\over {4m^2_N}}} G_T i\sigma_2
\right]_{\lambda'_N,\lambda_N}, \label{pion}
 \end{equation}
 where $\lambda'_N$, $\lambda_N$ are the recoil and target 
nucleon helicities, $G_V$  and $G_T$ are the vector and tensor 
 (helicity non-flip, and flip) $\rho NN$ couplings  and $R(t,s)$ is 
the $\rho$ meson, Regge propagator, 

\begin{equation}
R(s,t) = {1\over 2} \left(1 - e^{i\pi \alpha_\rho(t) } \right) 
\Gamma(1 - \alpha_\rho(t) ) \left(\alpha' s\right)^{\alpha_\rho(t)}, 
\end{equation}
with $\alpha_\rho(t) = 1 + \alpha'(t - m_\rho^2)$ and 
$\alpha' = 0.9\mbox{ GeV}^{-2}$. 

The $g_{X\rho\pi}$ coupling is normalized so that 
 the partial $\rho\pi$ width of the natural parity resonance is given by 

\begin{equation}
\Gamma_{X\to \rho\pi} = m_X {{ g^2_{X\rho\pi} } \over {32\pi^2}} \left(
{q\over m_X} \right)^{2J + 1}, 
\end{equation}
with $q$ being the on-shell breakup momentum, $q=\lambda(M_X,m_\pi,m_\rho)$. 
Finally, the differential cross section is then given by 

\begin{equation}
{{d\sigma} \over {dt}} = {{389\mu b GeV^2} \over {64\pi m_N^2p_L^2}} 
{1\over 2} \sum_{\lambda'_N,\lambda_N,\lambda_X} |A|^2. \label{cs}
\end{equation} 
For the $a_2$, the $\rho\pi$ partial width is known~\cite{PDG}, 
$\Gamma_{a2\rho\pi} = 0.7 \times 110\mbox{ MeV}$. 
Since we are primarily interested in  comparing the $a_2$ and the
$\pi_1$,  
 exotic meson we will  consider the production at $p_L=18\mbox{ GeV}$ 
 corresponding to the E852 data. 
 Prediction for the differential cross section for the $a_2$ 
 following from Eq.~(\ref{pion})
 is shown by the solid line in Fig.~1. 
 This can be compared with the $p_L=15\mbox{ GeV}$ charge exchange
 data from Ref.~\cite{a2pionold1}. The vector and tensor $\rho NN$ 
 couplings used are somewhat larger,  
 $G_V=5.9$, $G_T =27.7$ then the "standard" parameters 
 ( $G_V=2.3$, $G_T =18.4$)  
 and come from   what is referred to as the ``theoretical'' 
 parameterization
 in Ref.~\cite{WW}. One should note however, that even the
 ``standard'' parameters which come from the non-Regge $\rho$
 propagator after changing to the Regge propagators should be
 renormalized which enhances their values~\cite{Leon}. 
 The characteristic dip at $t\sim-0.5 \mbox{ GeV}^2$
 may be washed out by absorption (Regge+Pomeron) rescattering corrections. 
  Finally the low value of slope parameter $b_\rho = 2-3\mbox{GeV}^{-2}$ 
 in the residue function is typical to the $\rho$-exchange.

\begin{figure}
\centerline{\psfig{figure=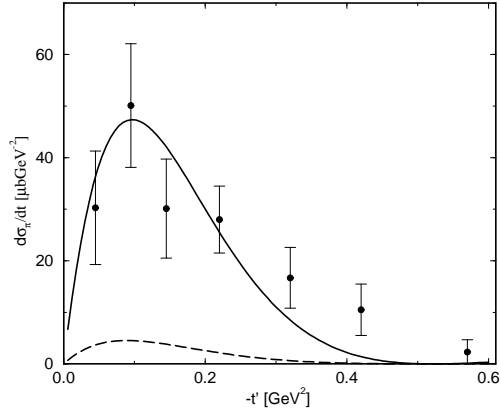,width=0.5\hsize}}
\smallskip
\caption{ Comparison of the $a_2(1320)$--regular meson (solid) and
  $\pi_1(1600)$--exotic meson (dashed) 
   production cross sections in reaction 
  $\pi^- p \to X \pi^+\pi^-\pi^0 n$ at $p_\pi = 18\mbox{ GeV}$. 
 The data correspond to $a_2$ production at $15\mbox{ GeV}$ } 
\label{fig1}
\end{figure}

 The E852 result indicates that the $\pi_1$ exotic wave found in the $\rho\pi$ 
 channel has the total width of $\Gamma_{\pi_1} \approx 170\mbox{ MeV}$ 
 and corresponds to about $5\%$ strength of the $a_2$. The $\pi_1$ 
  production is then expected to scale according to 
  
 \begin{equation}
 {{ \sigma_{\pi^- p \to \pi_1^0 n}  } \over 
  { \sigma_{\pi^- p \to a^0_2 n} } }  \approx  
 {{ \sigma_{\pi^- p \to \pi_1^- p}  } \over 
  { \sigma_{\pi^- p \to a^-_2 p} } }  \approx  
   {{ \sigma_{\pi N \to \pi_1 N \to \rho \pi N} }  \over 
   {  \sigma_{\pi N \to a_2 N \to \rho \pi N} } }
    {{ \Gamma_{a_2 \to \rho\pi} } \over  {\Gamma_{a_2} } }
     {{ \Gamma_{\pi_1} } \over  {\Gamma_{\pi_1 \to \rho\pi} }  }
      \approx 5\% \times 0.7 \times   {{ \Gamma_{\pi_1} } \over
  {\Gamma_{\pi_1 \to \rho\pi} }  }. \label{e852}
  \end{equation}
  The above relation  enables to determine the  
  $g_{\pi_1\rho\pi}$ coupling.
 We find that the $\Gamma_{\pi_1\to \rho\pi}$ partial width is
  approximately  $40\%$ of the total $\Gamma_{\pi_1} = 170\mbox{ GeV}$ 
   width. The $t$-dependence of the differential cross-section for 
 $\pi_1$ production is 
    shown by the dashed line in Fig.~1. 
 As expected it has a similar $t$-dependence to the 
    $a_2$ production, since under assumption that 
 the $\rho$-exchange dominates, the only difference comes from the 
 angular momentum barrier factors, 
    
\begin{equation}
    { {d\sigma_{\pi^- p \to \pi^0_1 N} }
 \over { d\sigma_{\pi^- p \to a^0_2 n}}} = {{3 g^2_{\pi_1\rho\pi}}\over
 {5g^2_{a_2\rho\pi}}} \left({ {2m^2_{a_2}} \over  { m_{a_2}^2 - t}}
 \right)^2 \approx 
4 {{3 g^2_{\pi_1\rho\pi}}\over
 {5g^2_{a_2\rho\pi}}} = 
  10\%, \label{pionratio}
 \end{equation}
 in agreement with the E852 result~\cite{rpi} scaled by the coupling to the
  $\rho\pi$ decay channel. Thus according the the E852
  measurement the exotic pion-production cross section is a small
  fraction of the $a_2$ production cross section. The corresponding
  couplings are calculated to be $g_{a2\rho\pi} \approx 75$ and $g_{\pi_1\rho\pi}
 \approx 16$. 

 We will now discuss charge exchange photoproduction. 
 The dominant production amplitude is now expected to come from
  one-pion-exchange with the photon acting, through VMD as a vector
  meson--$\rho$ beam.  On the basis of VMD we would expect, 
\begin{equation}
g^{VMD}_{\pi_1\gamma\pi} = {{\sqrt{4\pi\alpha}}\over {f_\rho}}C_f 
g_{\pi_1\rho\pi} \approx 0.04 g_{\pi_1\rho\pi} = 0.7\; .
\end{equation}
The radiative, $a_2$ width is known~\cite{PDG},
 $\Gamma_{a_2\gamma\pi} = 295\mbox{ keV}$ which gives
 $g_{a_2\gamma\pi} = 1.55$. The above VMD relation yields, 
\begin{equation}
g^{VMD}_{a_2\gamma\pi} = {{\sqrt{4\pi\alpha}}\over {f_\rho}}C_f
g_{a_2\rho\pi} \approx 0.04 g_{a_2\rho\pi} = 3\; .
\end{equation}
Here $C_f=1/\sqrt{2}=g_{X^\pm\rho^0\pi^\pm}/g_{X\rho\pi}$. 
 Thus for the radiative widths we expect the VMD to be accurate 
 within a factor of two. 
The charge-exchange, OPE photoproduction amplitude is given then by

\begin{equation}
A_\gamma =  m_X g_{X\rho\pi} \sqrt{ {{2J + 1}\over {4\pi}} }
 {{[\sigma^3]_{\lambda_X\lambda_\Gamma} } \over {\sqrt{2}}}
 g_{\pi NN}
[\sigma^1]_{\lambda'_N\lambda_N} 
\sqrt{{{-t'}\over {4m^2_N}}} R_\pi(t,s) e^{b_\pi t/2}
 \left( {{m_X^2 - t} \over {2m^2_X}}\right)^J \label{gamma},
 \end{equation}    
with

\begin{equation}
R(s,t) =  {1\over 2} \left(1 + e^{i\pi \alpha_\pi(t) } \right) 
\Gamma(- \alpha_\pi(t) ) \alpha' \left(\alpha' s\right)^{\alpha_\pi(t)} 
\end{equation}
being the $\pi$ Regge propagator with 
$\alpha_\rho(t) = \alpha'(t - m_\rho^2)$  and $g_{\pi NN} =
 35.5 \mbox{ GeV}$.

The existing data on $a_2$ photoproduction comes
 from two SLAC bubble chamber experiments one at the average photon energy of
 $E_\gamma = p_L = 4.8\mbox{ GeV}$~\cite{slac} and other at $E_\gamma=19\mbox{
   GeV}$~\cite{condo}. The corresponding total cross sections are
 measured 
to be  $\sigma_{\gamma p \to a^+_2 n} \approx 2.6\mu b$ and $0.3\mu b$ respectively. 
 The data for 
 differential cross section for the lower energy experiment is
  shown in Fig.~2 after rescaling by a factor of $2.6$ as explained in
 Ref.~\cite{condo}. This should be compared with the cross section 
 predicted from Eq.~(\ref{gamma}) (dashed line). The comparison is quite
 good, indicating, however need for some absorption 
 corrections~\cite{slac,asaf} 
 We will now compare the prediction for the exotic $\pi_1$
 photoproduction with that for the  $a_2$. We make the comparison
 for the photon energy, $E_\gamma = 8\mbox{ GeV}$ as proposed for the 
 Hall D experiments at JLab. This is shown in Fig.~2. The solid line
 is the prediction for the $a_2$ production and the shaded region 
  depicts the expected cross section for the $\pi_1$ using the 
  radiative width in the range $300\mbox{ keV} <
 \Gamma_{\pi_1\gamma\pi} < 400\mbox{ keV}$ which covers the typical
 range of radiative widths of ordinary mesons in this mass range and
  includes the VMD prediction, ($\Gamma^{VMD} =  
 m_X (g^{VMD}_{\pi_1\gamma\pi})^2 (q/m_X)^3/32\pi^2 = 360\mbox{ keV}$).

\begin{figure}
\centerline{\psfig{figure=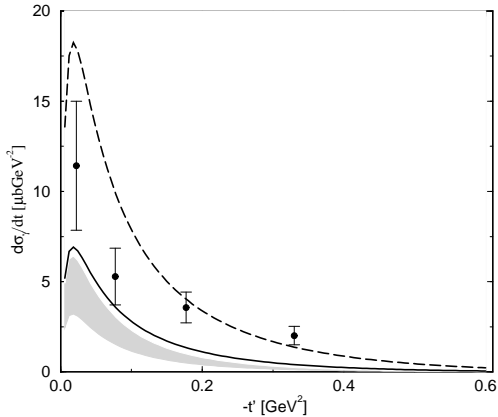,width=0.5\hsize}}
\smallskip
\caption{  Comparison of $a_2(1320)$ (solid) and exotic,
  $\pi_1(1600)$ (shaded region) 
    production cross sections in reaction 
  $\gamma p \to X^+ n$ at $p_\gamma = 8\mbox{ GeV}$. The data correspond to
   charge exchange $a_2$ production at $4.8\mbox{ GeV}$ and the
   theoretical prediction for $a_2$ at this energy is shown by the
   dashed line. The lowest (highest)  prediction for exotic meson production
 given by the shaded region corresponds to $\Gamma_{\pi_1\gamma\pi} = 200 (400) \mbox{ keV}$
   respectively. } 
\label{fig2}
\end{figure}

It is interesting to compare the predicted  ratio of the $\pi_1$ to
 $a_2$ photoproduction cross section. In analogy to Eq.~(\ref{pionratio}),
we get 

\begin{equation}
    { {d\sigma_{\gamma p \to \pi^+_1 n}}
 \over { d\sigma_{\gamma p \to a^+_2 n}}} = {{3 g^2_{\pi_1\gamma\pi}}\over
 {5g^2_{a_2\gamma\pi}}} { {m^2_{\pi_1}} \over {m_{a_2}^2}}
\left( {{ m_{\pi_1}^2 - t }\over {2 m_\pi^2}} \right)^2
 \left({ {2m^2_{a_2}} \over  { m_{a_2}^2 - t}}
 \right)^4 \approx 
 4  {{3 g^2_{\pi_1\gamma\pi}}\over
 {5g^2_{a_2\gamma\pi}}} { {m^2_{\pi_1}} \over {m_{a_2}^2}}
  \approx 50\%-100\%
 \label{gammaratio}
 \end{equation}
 and for small-t, 

\begin{equation}
 {     { d\sigma_{\gamma p \to \pi^+_1 n}/ 
   d\sigma_{\gamma p \to a^+_2 n} } \over 
  { d\sigma_{\pi^- p \to \pi^0_1 n}/ 
   d\sigma_{\pi^- p \to a^0_2 n} } } \approx 5-10
\end{equation}

{\it 3. Conclusions.} We have estimated the exotic meson photoproduction
 rate based on the existing data on hadronic production with pion beams. 
 From the E852 $\pi^- p \to X p \to \rho\pi p$ data it follows that 
 the exotic production is suppressed by
 roughly a factor of 10 as compared to the $a_2(1320)$ production. 
 We find, however, this is not the case in reactions with
 photon beams. Based on the rate estimate from the E852 data we
 conclude that in photoproduction the $\pi_1$ and $a_2$ production 
 should be comparable. 

There main reasons for enhancement of exotic production 
 is that the exotic meson 
 photocoupling  is not suppressed. This follows from simple kinematical 
constraints. The angular momentum barrier factors enhance the ratio of 
exotic to $a_2$ photocouplings over the respective ratio of hadronic, 
 $\rho\pi$ couplings. Assuming similar hadronic $X\to \rho\pi$ and  
 radiative, $X\to\gamma\pi$ decay widths, 

\begin{equation}
 {  {g^2_{\pi_1\gamma\pi}/g^2_{a_2\gamma\pi}} \over
    {g^2_{\pi_1\rho\pi}/g^2_{a_2\rho\pi} } } 
 = \left( {{\lambda(m_{a_2},\rho,\pi)/\lambda(m_{\pi_1},\rho,\pi)}
      \over{\lambda(m_{a_2},0,\pi)/\lambda(m_{\pi_1},0,\pi)}}
 \right)^2 \approx 4\; .
\end{equation}
Secondly since the small-t behavior is 
 determined by the helicity structure of the production amplitude, both the
 exotic and $a_2$ are enhanced at low-t in photoproduction via OPE 
 as compared to hadronic production via $\rho$-exchange. 
 This is due  to the 
 vector nature of the photon which transfers its helicity to the
 produced meson. This is not the case for the pion beam, where helicity of the
 produced meson has to be transferred from the exchanged (vector) meson,
 and thus making the amplitude suppressed with $\sqrt{-t'}$. 
 These findings are important for exotic mesons searches in future
 photoproduction experiments in Hall D at the JLab.

{\it 4. Acknowledgment.} We would like to thank Alex Dzierba and Dennis
Weygand for stimulating discussions. This work was supported by DOE
grant under contract  DE-FG02-87ER40365.

\vglue 0.4cm

\end{document}